\newcommand{\lumifb}{\mbox{fb$^{-1}$}}				
\newcommand{\babar}{\mbox{\sl B\hspace{-0.4em} {\small\sl A}\hspace{-0.37em} \sl B\hspace{-0.4em} {\small\sl A\hspace{-0.02em}R}}}
\newcommand{\etal}{\textit{et al.}}

\newcommand{\Bhh}{\ensuremath{B^0_{(s)} \to h^+ h'^{-}}}

\newcommand{\Bdpipi}{\ensuremath{B^0 \to \pi^+ \pi^-}}
\newcommand{\BdKpi}{\ensuremath{B^0 \to K^+ \pi^-}}
\newcommand{\BdKK}{\ensuremath{B^0 \to K^+ K^-}}

\newcommand{\Bspipi}{\ensuremath{B^0_s \to \pi^+ \pi^-}}

\newcommand{\BsKK}{\ensuremath{B^0_s \to K^+ K^-}}

\newcommand{\BR}{\ensuremath{\mathcal{B}}}

\documentclass{PoS}

\title{Updated measurements of hadronic $B$ decays at CDF}

\ShortTitle{Updated measurements of hadronic $B$ decays at CDF}

\author{\speaker{Michael J. Morello}\thanks{On behalf of the CDF Collaboration.}\\
        Scuola Normale Superiore di Pisa and INFN Sezione di Pisa\\
        E-mail: \email{michael.morello@pi.infn.it}}

\abstract{
The CDF experiment at the Tevatron $p\bar{p}$ collider established that extensive and detailed exploration of the $b$--quark 
dynamics is possible in hadron collisions, with results competitive and supplementary to those from $e^+e^-$ colliders. 
This provides a rich, and highly rewarding program that has currently reached full maturity. 
In the following I report some recent results on hadronic decays: 
the evidence for the charmless annihilation decay mode \Bspipi, and 
the first reconstruction in hadron collisions of the suppressed decays $B^- \to D(\to K^+\pi^-)K^-$ and $B^- \to D(\to K^+\pi^-)\pi^-$.
}

\FullConference{The 2011 Europhysics Conference on High Energy Physics, EPS-HEP 2011,\\
		July 21-27, 2011\\
		Grenoble, Rh\^one-Alpes, France}

\begin{document}

\section{Evidence for the charmless annihilation decay mode \Bspipi}

Two-body non-leptonic charmless decays of $b$--hadrons are very 
widely studied processes in flavor physics. 
The variety of open channels involving similar final states provides 
useful experimental information to improve the accuracy of 
effective models of strong interaction dynamics. 
Some decays receive contributions from higher-order (`penguin') transitions, 
and are therefore sensitive to the possible presence of new physics in internal loops.

The \Bspipi\  and  \BdKK\ decay modes have a special status, in that
all quarks in the final state are different from those in the initial state.  
This limits the possible diagrams that can contribute to these decay
to penguin-annihilation (PA) and $W$-exchange (E) topologies. 
They have never been observed, and the best upper limits come from \cite{Abe:2006xs,Aaltonen:2008hg}.
These amplitudes are difficult to predict within the current
phenomenological models, and are often neglected in calculations of
decays where they are not the only contributors, using generic
argument of smallness. These diagrams may carry different CP-violating
and CP-conserving phases with respect to leading diagrams; the lack of
knowledge of their size therefore introduces unavoidable uncertainties
in the conclusions drawn from the analyses of other well-studied decays, like \Bdpipi\ and \BsKK. 
A measurement of  branching fraction of both \Bspipi\ and \BdKK\  would be particularly useful, as it allow a 
better determination of the strength of PA and E amplitudes. 

\begin{figure}[ht]
\begin{center}
\includegraphics[width=.45\textwidth]{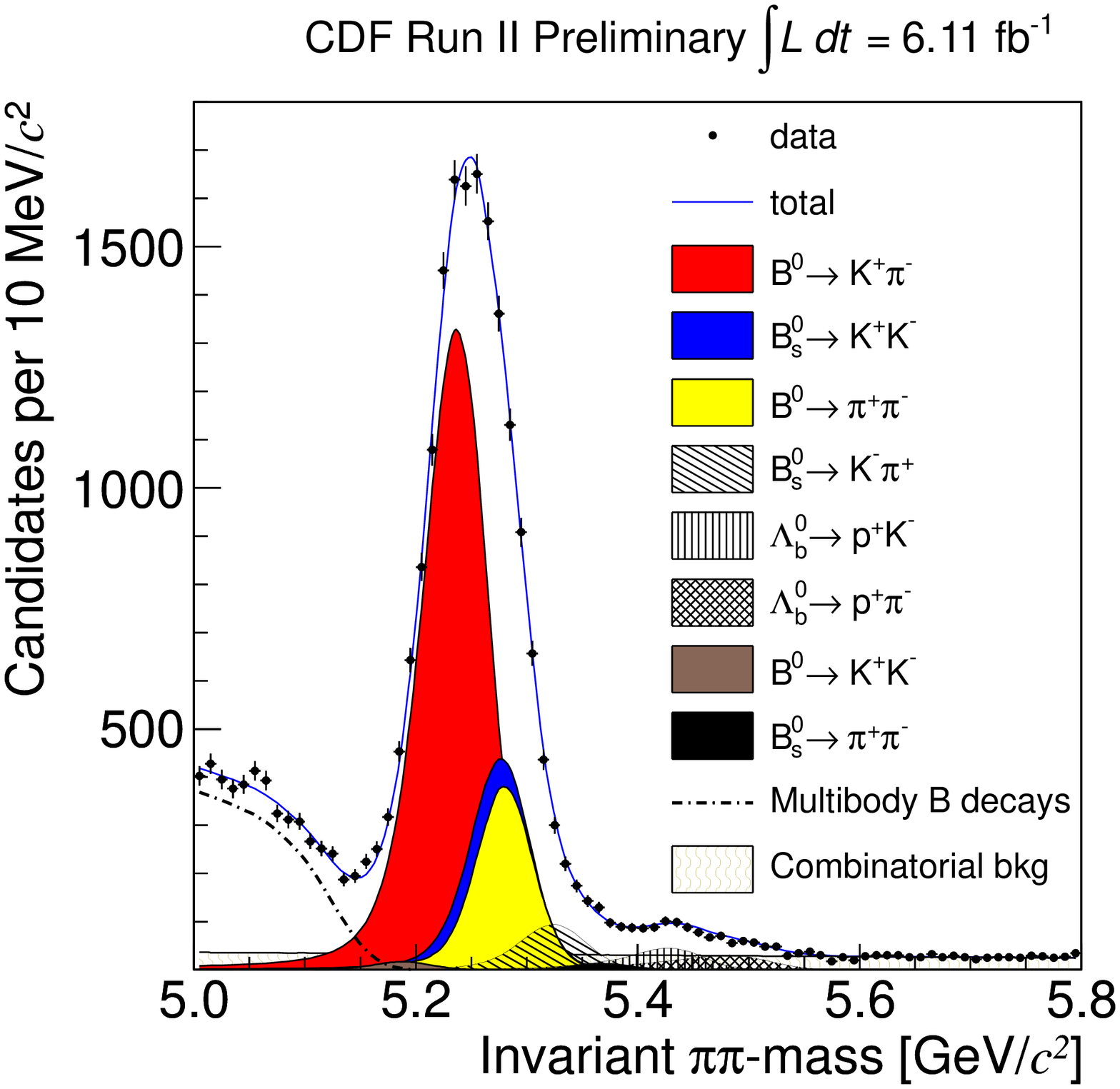} 
\hspace{1cm}
\includegraphics[width=.45\textwidth]{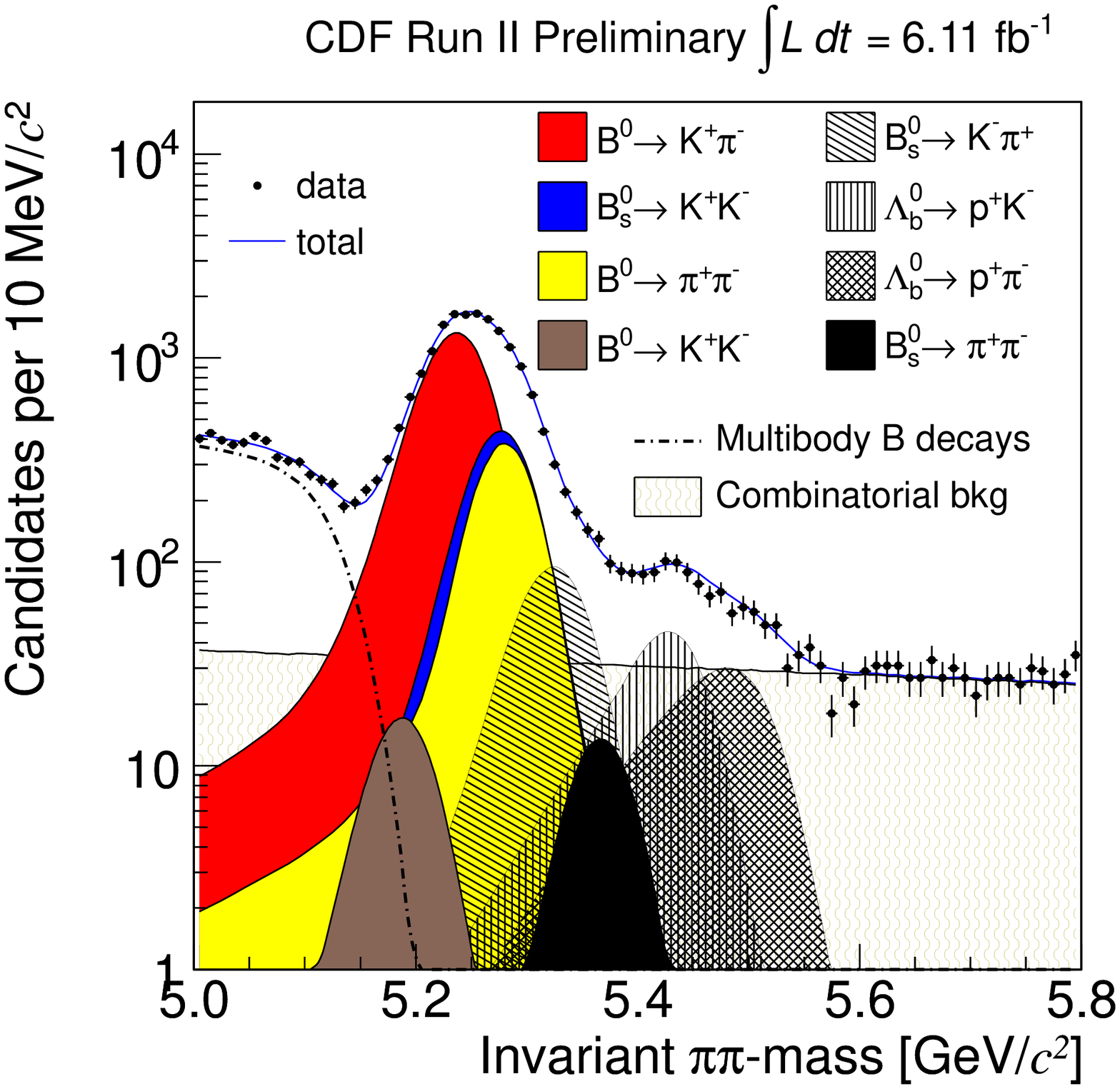} 
\end{center}
\caption{Invariant mass distribution of reconstructed \Bhh\ candidates (linear scale on the left and logarithmic scale on the right). The charged pion mass is assigned to both tracks.
The sum of the fitted distributions and the individual components of signal and background are overlaid on the data distribution. 
}
\label{fig1} 
\end{figure}

CDF recently searched for both \Bspipi\ and \BdKK, using a data sample of 6\lumifb\ of 
integrated luminosity. An extended unbinned likelihood fit,
incorporating kinematic (invariant mass and momenta) 
and particle identification (dE/dx) information, is used to determine the
fraction of each individual \Bhh\ (where $h$ is for a pion or kaon) mode in the sample.
The fit projection on the invariant $\pi\pi$-mass is reported in fig.~\ref{fig1}.
A 3.7$\sigma$ significant signal is observed for the \Bspipi\ mode, 
while a deviation at the $2\sigma$ level from no signal hypothesis is reported for the \BdKK. 
The measured branching ratios are  $\BR(\Bspipi)=(0.57\pm0.15\pm0.10)\times 10^{-6}$ and 
$\BR(\BdKK)=(0.23\pm0.10\pm0.10)\times 10^{-6}$~\cite{bspipi_cdf}. 
A 90\% confidence interval for the \BdKK\ mode is also  reported  $[0.05,0.46]\times10^{-6}$~\cite{bspipi_cdf}.
The result for the \Bspipi\ mode is consistent 
with the previous upper limit ($< 1.2 \times 10^{-6}$ at 90\%~C.L.), based on a subsample of the current data 
\cite{Aaltonen:2008hg} and it is in agreement with the very recent result from LHCb~\cite{bspipi_lhcb}. 
This agrees with the predictions in Ref.~\cite{bspipi_pqcd}  within the pQCD approach.
It is higher than most other predictions \cite{Beneke:2003zv,Sun:2002rn,Cheng:2009mu}.
The present measurement of $\BR(\BdKK)$ 
supersedes the previous limit~\cite{Aaltonen:2008hg}. The central value is in agreement 
with other existing measurements~\cite{Aubert:2006fha,Abe:2006xs,bspipi_lhcb}, 
and with theoretical predictions~\cite{Beneke:2003zv,Cheng:2009mu}.

\section{\boldmath{$\gamma$} from \boldmath{$B \to D K$} decays}

Conventionally, CP violating observables are written in terms of the angles $\alpha$, $\beta$ and $\gamma$ of 
the ``Unitarity Triangle", obtained from one of the unitarity conditions of the CKM matrix. 
While the resolution on $\alpha$ and $\beta$ reached a good level of precision, the measurement of $\gamma$ is still limited 
by the smallness of the branching ratios involved in the processes. 
Among the various methods for the $\gamma$ measurement, those which make use of the tree-level $B^- \to D^0 K^-$ 
decays have the smallest theoretical uncertainties. 
In fact $\gamma$ appears as the relative 
weak phase between two amplitudes, the favored $b \to c \bar{u} s$ transition of the $B^- \to D^0 K^-$, whose amplitude is 
proportional to $V_{cb} V_{us}$, and the color-suppressed $b \to u \bar{c} s$ transition of the $B^- \to \overline{D}^0 K^-$, 
whose amplitude is proportional to $V_{ub} V_{cs}$.  
The interference between $D^0$ and $\overline{D}^0$, decaying into the same final state, leads to measurable CP-violating 
effects, from which $\gamma$ can be extracted. The effects can be also enhanced choosing the interfering amplitudes of the same order of magnitude.
All methods require no tagging or time-dependent measurements, and many of them only involve charged particles in 
the final state. 
%
\begin{figure}[ht]
\begin{center}
\includegraphics[width=.45\textwidth]{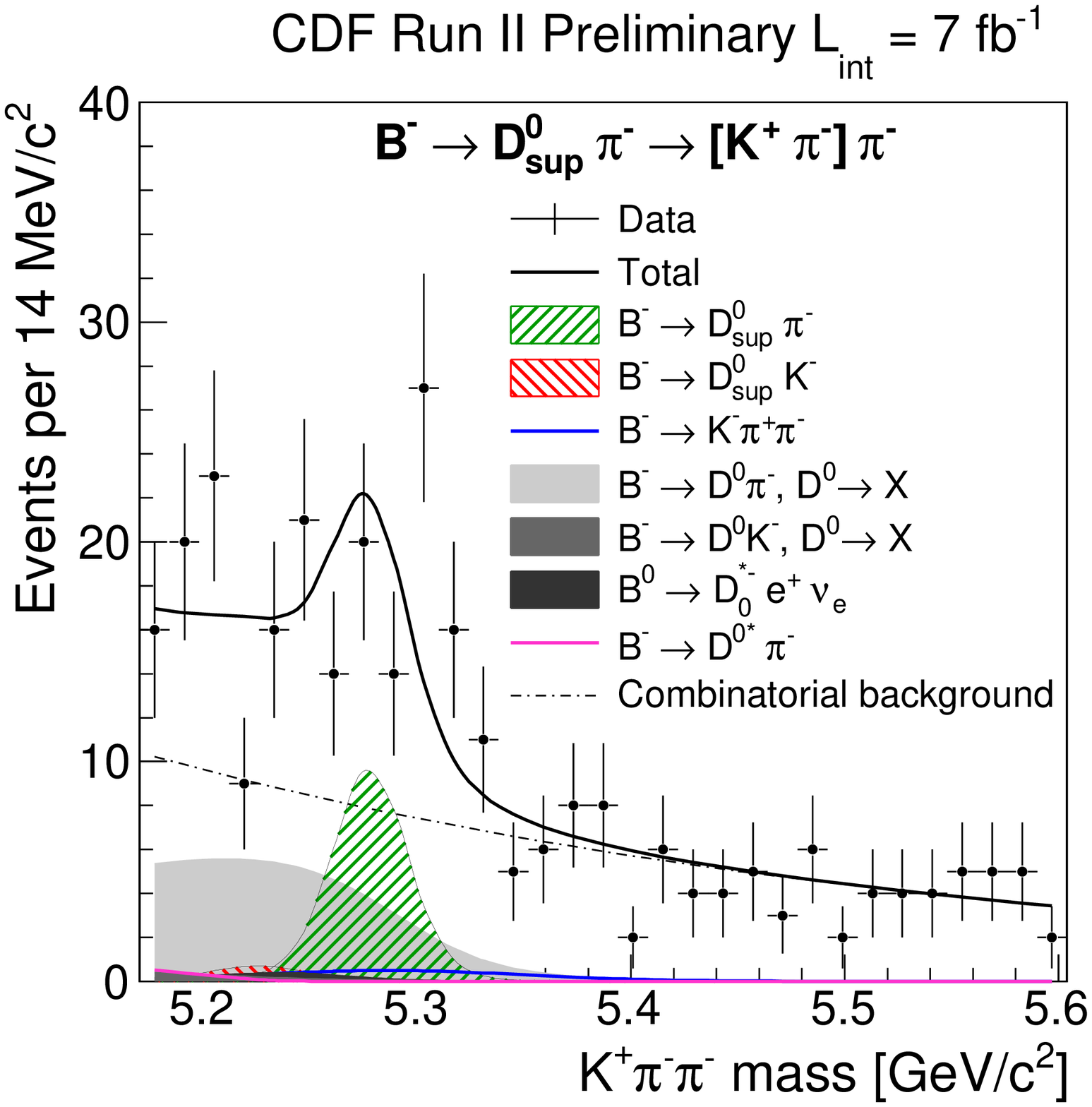} 
\hspace{1cm}
\includegraphics[width=.45\textwidth]{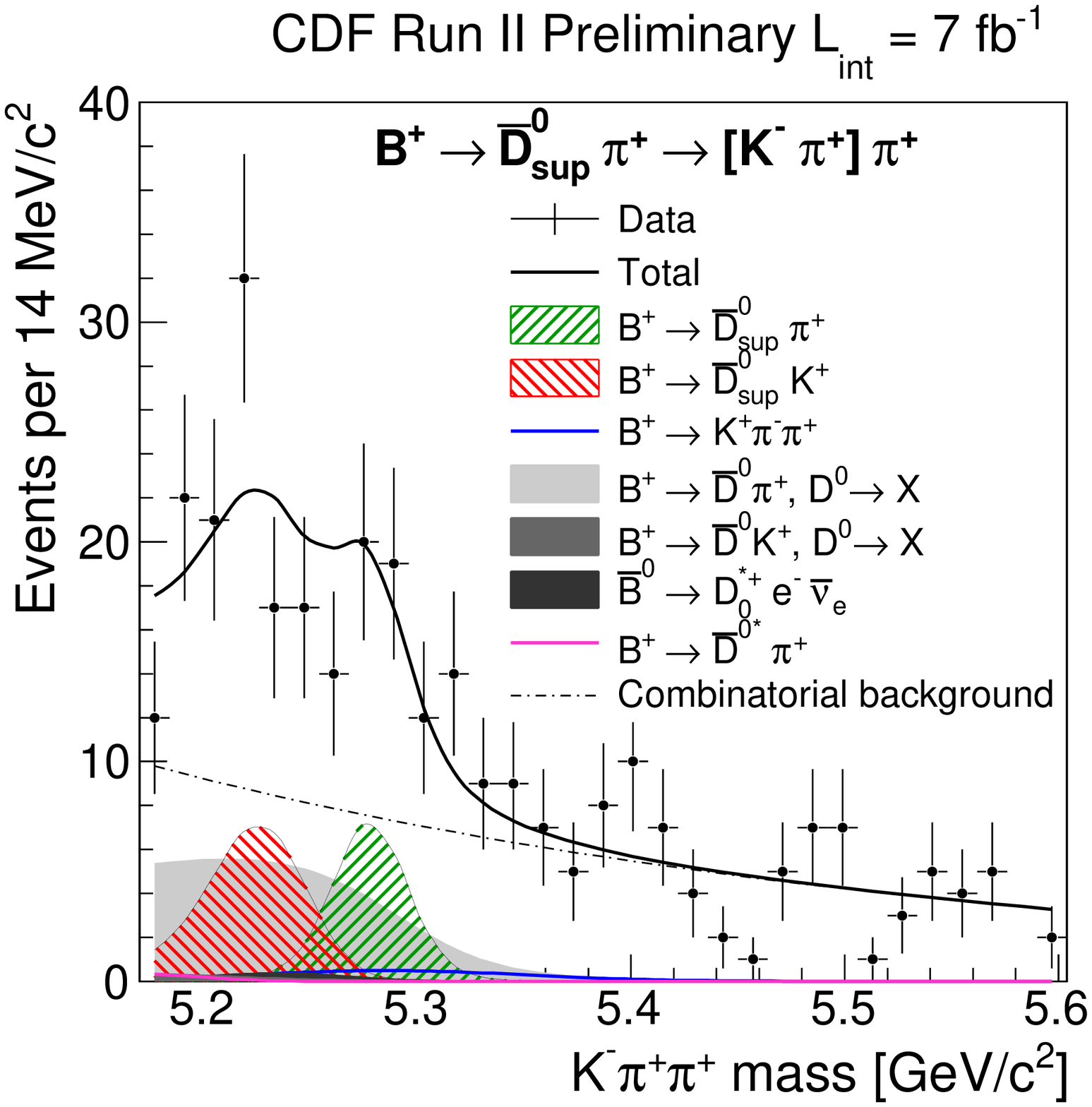} 
\end{center}
\caption{Invariant mass distributions of $B^{\pm} \to Dh^{\pm}$ for the suppressed mode (bottom meson on the left and antibottom on the right).
The pion mass is assigned to the charged track from the B candidate decay vertex. The projections of the likelihood fit are overlaid.
}
\label{fig2} 
\end{figure}

In a data sample of about 7~\lumifb\ CDF reports the first reconstruction in hadron collisions of the suppressed decays 
$B^- \to D(\to K^+ \pi^-)K^-$ and $B^- \to D(\to K^+ \pi^-)\pi^-$, which are the main ingredient of 
the the ADS method~\cite{ref:ads1}.
Also in this case an extended unbinned likelihood fit,
incorporating kinematic (invariant mass) 
and particle identification (dE/dx) information, is used to determine the
fraction of each individual modes. The fit projection on the invariant $K\pi\pi$-mass is reported in fig.~\ref{fig2}.
CDF measures the following asymmetries:
$A_{ADS}(K)  =   -0.82 \pm 0.44\mbox{(stat)} \pm 0.09\mbox{(syst)}$ and 
$A_{ADS}(\pi) =   0.13 \pm 0.25\mbox{(stat)} \pm 0.02\mbox{(syst)}$~\cite{cdf_ads},
and for the ratios of doubly Cabibbo suppressed mode to flavor eigenstate CDF finds 
$R_{ADS}(K)  =  [22.0 \pm 8.6\mbox{(stat)} \pm 2.6\mbox{(syst)}] \times 10^{-3}$ and 
$R_{ADS}(\pi)  =  [2.8 \pm 0.7\mbox{(stat)} \pm 0.4\mbox{(syst)}] \times 10^{-3}$~\cite{cdf_ads}.
The results are in agreement with existing measurements 
performed at $\Upsilon$(4S) resonance~\cite{ads_bfactories} and very recently at LHCb~\cite{ads_lhcb}. 

\section{Conclusions}

CDF experiment at the Tevatron keeps providing excellent results in the exploration of
Heavy Flavor Physics,  owing to CP-symmetric initial states in $\sqrt{s}=1.96$ TeV $p\bar{p}$ collisions, 
large event samples collected by well-understood detector, and mature analysis techniques.
In summary, this short write-up reports on the first evidence for the charmless annihilation decay mode \Bspipi, 
an updated upper limit for the \BdKK\ mode,  and on the first reconstruction in hadron collisions 
of the suppressed decays $B^- \to D(\to K^+\pi^-)K^-$ and $B^- \to D(\to K^+\pi^-)\pi^-$.

\end{document}